%Paper: hep-ph/9507321
%From: forte@surya11.cern.ch (Stefano Forte)
%Date: Mon, 17 Jul 95 19:53:59 +0200

%%%%%%%%%%%%%%%%%%%%%%%%%%%%%%% v3.tex %%%%%%%%%%%%%%%%%%%%%%%%%%%%%%%%%%
%%%%%%%              Momentum Conservation at Small x              %%%%%%
%%%%%%%                 R. D. Ball and S. Forte                    %%%%%%
%%%%%%%      in plain TeX with the harvmac macro package           %%%%%%
%%%%%%%%%%%%%%%%%%%%%%%%%%%%%%%%%%%%%%%%%%%%%%%%%%%%%%%%%%%%%%%%%%%%%%%%%

\input harvmac

%\draftmode
\noblackbox

\pageno=0\nopagenumbers\tolerance=10000\hfuzz=5pt
\line{\hfill CERN-TH/95-198}
\line{\hfill \tt hep-ph/9507321}
\vskip 24pt
\centerline{\bf MOMENTUM CONSERVATION AT SMALL $x$}
\vskip 36pt\centerline{Richard~D.~Ball\footnote{*}{On leave
 from a Royal Society University Research Fellowship.}
 and Stefano~Forte\footnote{\dag}{On leave
 from INFN, Sezione di Torino, Turin, Italy.}}
\vskip 12pt
\centerline{\it Theory Division, CERN,}
\centerline{\it CH-1211 Gen\`eve 23, Switzerland.}
\vskip 36pt
{\centerline{\bf Abstract}
\medskip\narrower
\ninepoint\baselineskip=9pt plus 2pt minus 1pt
\lineskiplimit=1pt \lineskip=2pt
We discuss how momentum conservation is implemented in perturbative
computations based on expansions of anomalous dimensions appropriate
at small $x$. We show that for any given choice of $F_2$ coefficient
functions there always exists a factorization scheme where the gluon
is defined in such a way that momentum is conserved at next to leading
order.
\smallskip}
\vskip 20pt \centerline{Submitted to: {\it Physics Letters B}}
\vskip 48pt
\line{CERN-TH/95-198\hfill}
%\line{\tt hep-ph/9507321\hfill}
\line{July 1995\hfill}

\vfill\eject \footline={\hss\tenrm\folio\hss}

%#############################################################################%

\def\frac#1#2{{{#1}\over {#2}}}
\def\half{\hbox{${1\over 2}$}}

\def\smallfrac#1#2{\hbox{${{#1}\over {#2}}$}}

\def\MS{\hbox{$\overline{\rm MS}$}}
\def\DIS{\hbox{DIS}}
\def\SDIS{\hbox{SDIS}}
\def\GDIS{\hbox{GDIS}}

\def\QDIS{\hbox{Q$_0$DIS}}
\def\g{\gamma}\def\a{\alpha}
\catcode`@=11 %This allows us to modify plain macros
\def\slash#1{\mathord{\mathpalette\c@ncel#1}}
 \def\c@ncel#1#2{\ooalign{$\hfil#1\mkern1mu/\hfil$\crcr$#1#2$}}
\def\lsim{\mathrel{\mathpalette\@versim<}}
\def\gsim{\mathrel{\mathpalette\@versim>}}
 \def\@versim#1#2{\lower0.2ex\vbox{\baselineskip\z@skip\lineskip\z@skip
       \lineskiplimit\z@\ialign{$\m@th#1\hfil##$\crcr#2\crcr\sim\crcr}}}
\catcode`@=12 %at signs are no longer letters

\def\NP{{\it Nucl.~Phys.~}}

\def\PL{{\it Phys.~Lett.~}}

\def\ZP{{\it Zeit.~Phys.~}}

\def\vol#1{{\bf #1}}\def\vyp#1#2#3{\vol{#1} (#2) #3}

%$$$$$$$$$$$$$$$$$$$$$$$$$$$$$$$$$$$$$$$$$$$$$$$$$$$$$$$$$$$$$$$$$$$$$$$$$$$$%

An accurate determination of the $x$ and $Q^2$ dependence of structure
functions at small $x$ requires solution of next to leading order (NLO)
evolution equations appropriate to this
kinematical region~\ref\Alf{R.~D.~Ball and S.~Forte, CERN
preprints CERN-TH/95-132,
{\tt hep-ph/9505388}; CERN-TH/95-148, {\tt hep-ph/9506233}.}.
The perturbative expansion of the anomalous dimensions
or splitting functions which govern
perturbative evolution can then be reorganized in order to keep into
account the
 presence of two large scales in the problem ($Q^2$ and
$s=(1-x)Q^2/x$)~\ref\Summing{R.D.~Ball and S.~Forte,
\PL\vyp{B351}{1995}{313}.}.
The results of this procedure must be consistent with the constraints
imposed by  conservation laws. Energy-momentum
conservation is particularly subtle  in this context because
energy is now
one of the parameters which organize the perturbative expansion: therefore,
it can only be imposed by an appropriate definition of the
infinite set of contributions
which are summed up in the anomalous dimensions. The problem of
momentum conservation can thus be tackled only after these contributions have
been defined in the most general way, in particular by correctly accounting
for the freedom of choosing a factorization scheme.
Here we will show that momentum conservation can always be implemented in NLO
computations performed in expansion schemes appropriate to small $x$
by a judicious choice of factorization scheme. After defining
the appropriate small-$x$ expansions, we will construct the most
general transition functions which perform a scheme change. We will then show
that given a matrix of anomalous dimensions and a factorization scheme,
specified by the coefficient functions which relate $F_2$ to parton
distributions (such
as, for instance, \MS\ or DIS) there is still enough freedom to
perform a further scheme change within the given scheme (i.e. without
changing the coefficient functions) such that momentum conservation is then
obtained consistently at NLO.

A determination of the evolution of parton distributions by solution
of the renormalization group equations corresponds to summing
leading (and subleading) logs of the form $\a_s^p (\log Q^2)^q
(\log {1\over x})^r$. In the usual loop expansion, appropriate to the
Bjorken limit, the leading logs are those of $Q^2$. Thus at leading order
all terms with $p=q$ are summed, at NLO those with $q<p\leq 2q$, and
so forth; it then turns out that both at LO and NLO (and, in fact, at any
order) $0\le r\le p$.
At small $x$, however, logs of ${1\over x}$ should also be considered
leading, and the perturbative
expansion reorganized accordingly. It is for example
possible~\Summing\ to define an expansion scheme appropriate to the
Regge limit where the roles
of $\ln{1\over x}$ and $\ln Q^2$ are interchanged, so that at LO
all logs with $p=r$ are summed, while $1\le q\le p$ (the small $x$ expansion).
A scheme where the two logs are treated on the same footing
(double leading expansion) can also be defined,
in which at LO any power of $\a_s$ is
accompanied by either of the two logs (or $1\le q\le p$, $0\le r\le p$,
$1\le p\le q+r$), as well as a number of intermediate schemes.
In the sequel we will consider specifically the small $x$ expansion,
which is a theoretically interesting limiting case, and the double
leading expansion, which is most interesting for phenomenology in the HERA
region~\Alf.

The constraint imposed by  momentum  conservation on anomalous dimensions is
a particular case of the general requirement that anomalous dimensions of
conserved (or partially conserved) operators must vanish, which in turn is
a consequence of the renormalization group equations. This constraint
 takes the form
\eqn\momcon{\g_{qg}(1,\a)+\g_{gg}(1,\a)=0,\qquad
\g_{qq}(1,\a)+\g_{gq}(1,\a)=0,}
where the anomalous dimensions
are moments of the splittings
functions,
$\gamma_{ij}(N,\alpha)=
\int_0^1 x^N P_{ij}(x;t)$, and depend on $t=\ln (Q^2/\Lambda^2)$ through the
running of the coupling $\alpha$. Momentum conservation [eq.~\momcon]
is imposed order by order in the usual loop expansion
in powers of $\a$ of the anomalous dimensions,
by suitable choice of normalization of the quark and gluon distributions.
If, however, the order $\a^k$ contribution to
$\gamma_N^{ij}$ is further expanded in powers of $N$ the single terms of this
expansion  will not, of course, satisfy eq.~\momcon. Now, the
various expansion schemes alluded above are obtained precisely by
performing such expansions, and then
including at each order a suitable subset of terms, which then will not
conserve momentum automatically.

When evolution in the small $x$ region is approached
by choosing an appropriate expansion scheme, momentum
conservation should be imposed order by order, just as
it is in the loop expansion.\foot{Alternatively, momentum conservation
could always be imposed at each order in $\alpha$ by including an
Ansatz for the (yet unknown) terms
which are formally sub-subleading in the small $x$ expansions: several
proposals of this kind are discussed in ref.~\ref\EHW{R.K.~Ellis,
F.~Hautmann and B.R.~Webber,\PL\vyp{B348}{1995}{582}.}.} That this
is a priori non-trivial is clear from the observation that the leading order
anomalous dimensions in the small $x$
expansion described above violate eq.~\momcon. Just like
in the usual expansion, however, eq.~\momcon\ only holds if
parton distributions
are defined appropriately, i.e. for suitable choices of the factorization
scheme. Before we discuss the implementation of momentum conservation
we must therefore discuss changes of factorization scheme
within various small $x$ expansions.

Changing the factorization scheme amounts~\ref\FP{W.~Furmanski and
R.~Petronzio,
\ZP\vyp{C11}{1982}{293}.} to a
redefinition of the singlet parton densities $f(N,t)\equiv
\bigl({q\atop g}\bigr)$, where $q\equiv\sum_i (q_i+\bar q_i)$. Letting
$f\to f'=Uf$, the naive partonic interpretation of the
parton densities will be
maintained if we always assume that $U(N,\a)=1+O(\a)$. The renormalization
group equation $\frac{d}{dt}f=\g f$ then remains unchanged
provided $\g\to\g'$, where
\eqn\gamsch{\g' = U\g U\inv +\big(\smallfrac{d}{dt}U\big)U\inv.}
Because $U$ only depends on $t$ through $\alpha$
\eqn\deru{\smallfrac{d}{dt}U(N,\a)
\equiv\beta(\a)\smallfrac{\partial}{\partial\a}U(N,\a),}
where $\beta(\a)$ is the beta function. Since
$\beta(\a)=-\beta_0\a^2+O(\a^3)$ it follows that the second term of
\gamsch\ is subleading compared to the first.

We must now choose a specific expansion scheme. We will eventually
prove our result in the physically relevant double leading expansion; however,
we consider first the small $x$ expansion,
since this will allow us to present the general structure of our results
in a somewhat simpler setting.
In the small-$x$ scheme the singlet anomalous dimension are given by
\eqn\singad{
\eqalign{
\g&\equiv \g^{s}+\g^{ss}+\cdots;\cr
\g^{s}(N,\a)&\equiv\sum_{n=n_0}^\infty \g^{s}_n(\a/N)^n\cr
\g^{ss}(N,\a)&\equiv
\a\sum_{n=n_0-1}^\infty \g^{ss}_{n+1}(\a/N)^{n},\cr}}
where $\g^{s}(N,\a)$
sums the leading singularities, $\g^{ss}(N,\a)$ the subleading
singularities, and so on. The one loop contributions to
$\g^{s}$ and $\g^{ss}$ [i.e. the contributions with
$n=1$ in eq.~\singad] will in general violate the condition~\momcon;
they are scheme independent
and thus lead inevitably to momentum nonconservation in the usual
small $x$ expansion if $n_0=1$. We will hence consider here
a truncated small $x$ expansion with $n_0=2$, i.e. with these
one loop contributions suppressed. The results then found
will coincide with those of the usual small $x$ scheme in the $x\to 0$ limit;
we will use them to prove that momentum conservation holds in the
double leading expansion.

At NLO in a typical factorization scheme
such as the \QDIS\ scheme \ref\ciaf{M. Ciafaloni,
CERN preprint CERN-TH/95-119, {\tt hep-ph/9507307}.}
 (or the \MS\ and DIS
schemes~\ref\CHad{ S.~Catani \& F.~Hautmann, \PL\vyp{B315}{1993}{157},
                  \NP\vyp{B427}{1994}{475}.}) $\g$ is then of the form
\eqn\nload{\g = \pmatrix{r\g_q & \g_q \cr
r\g_g+\hat\g & \g_g\cr}
= \pmatrix{0&0\cr r\g_g^{s} & \g_g^{s}\cr}
+ \pmatrix{r\g_q^{ss} &\g_q^{ss}\cr
r\g_g^{ss}+\hat\g_g &\g_g^{ss}\cr}+\cdots,}
where $r\equiv C_F/C_A$ is the colour-charge factor.
It is important for what follows to notice that both of the
quark anomalous dimensions vanish at leading order; also, both
the gluon anomalous dimensions
at leading order, and the quark ones at subleading order obey a
colour-charge relation.\foot{Notice that this
relation is however not satisfied by the one-loop term $\g^{ss}_1$,
which has been suppressed.}  At subleading order in the gluon
sector the anomalous dimensions $\g^{ss}_{gq}$ and $\g^{ss}_{gg}$
are as yet unknown beyond two loops so we introduce
$\hat\g_g\equiv\g^{ss}_{gq}-r\g^{ss}_{gg}$ to allow for possible
violations of the colour-charge relation at NLO in the gluon channel.

Consider now parton distributions
normalized so as to satisfy the
momentum sum rule at some scale $t=0$: $q(1,0)+g(1,0)=1$. Momentum is
then conserved in the evolution of
these distributions to the scale $t$ if the two conditions \momcon\
are satisfied. With  anomalous
dimension of the form \nload, the momentum sum rule will be violated at
LO (and thus in a scheme invariant way)
since $\g_g^{s}(1,\a)\neq 0$. However,  the quark
distribution only evolves at NLO, and the gluon can only be
directly observed at NLO (through measurement of $F_L$, say);
therefore, this LO violation
has no physical effects. Momentum conservation starts thus being physically
relevant at NLO,
where it imposes a condition relating the LO and NLO
components of the anomalous dimension, as well as  the
colour-charge relation in the gluonic sector:
\eqn\msx
{(\g_q^{ss})_n+(\g_g^{s})_n+(\g_g^{ss})_n=0,\qquad(\hat\g_g)_n=0,}
where $\g^s_g\equiv \sum (\g_g^{s})_n(\a/N)^n$, etc.
The conditions \msx\ will not in general be satisfied in a given
generic scheme; however in what follows we will show that it
is always possible to find factorization schemes such that both
conditions are satisfied (for $n>1$), and thus in which momentum is
conserved.

Consider first a scheme change $U$ which is LO in the small-$x$ expansion,
i.e. $U\equiv 1+\sum_1^\infty U_n(\a/N)^n$.
The most general form of U which retains the
identification of $F_2$ with the quark density at LO in this
expansion is\foot{We
neglect LO transformations proportional to the unit matrix, since they
modify the LO relation between $F_2$ and $q$.}
\eqn\udef{U\equiv\pmatrix{1&\bar u\cr\tilde u&u\cr},}
where $u\equiv1+\sum_1^\infty u_n(\a/N)^n$ while $\tilde u\equiv
\sum_1^\infty \tilde u_n(\a/N)^n$ and similarly for $\bar u$.
Substitution in \gamsch\ gives to LO
\eqn\gamprime{\g'= \g_g(u-\bar u\tilde u)\inv
\pmatrix{\bar u(ru-\tilde u)&\bar u(1-r\bar u)\cr
u(ru-\tilde u)&u(1-r\bar u)\cr}
+O(\a).}
If we now insist that the leading order anomalous dimension
is to remain unchanged, we must
choose~\nref\sdis{S.~Catani, Florence preprint DFF 226/5/95,
{\tt hep-ph/9506357}.}
\eqn\colchg{\bar u=0,\qquad\tilde u=r(u-1).}
A LO scheme change thus amounts essentially
to a redefinition of the
gluon normalization by the function $u$.
The NLO anomalous dimensions are then correspondingly modified according to
\eqn\gamprcc{\g'=\g +(u\inv-1)\g_q
\pmatrix{r&1\cr-r^2&-r\cr}
+(u-1)\hat\g_g\pmatrix{0&0\cr 1&0\cr}
+\smallfrac{d}{dt}\ln u\pmatrix{0&0\cr r&1\cr},}
so that in particular $\gamma^{qg}\equiv\g_q\to\g_q/u$.
Notice that the colour-charge relation is automatically preserved not
only  in the LO gluon
sector, but also in the NLO quark sector.

The quark distribution is left unaffected by the LO
transformations considered so far. However, we may still perform
 a NLO redefinition $f'\to f''=(1+V)f'$,
where
\eqn\vexp{
V\equiv\a\sum_0^\infty V_{n+1}(\a/N)^n.}
This induces a
corresponding change in the anomalous dimension
\eqn\gampp{\g'\to\g''=\g'+[V,\g']+O(\a^2).}
Writing
\eqn\vdef{V\equiv\pmatrix{\tilde v&v\cr\tilde w&w\cr},}
and keeping only LO and NLO terms, we find
\eqn\gpp{\g''=\g'+\g_g'\pmatrix{rv&v\cr
(rw-\tilde w-r\tilde v)&-rv\cr}+O(\a^2).}
Combining this NLO transformation with the LO transformation
$U$, we thus have altogether at NLO
\eqn\twotfmns{\eqalign{\g_{qq}''&=\g_{qq}
+r(u\inv-1)\g_q^{ss}+rv\g_g^{s},\cr
\g_{qg}''&=\g_{qg}+(u\inv-1)\g_q^{ss}+v\g_g^{s},\cr
\g_{gq}''&=\g_{gq}+(u-1)(\hat\g_g+r\hat\g_q)
-r^2(u\inv-1)\g_q^{ss}+(rw-\tilde w-r\tilde v)\g_g^{s}
+r\smallfrac{d}{dt}\ln u,\cr
\g_{gg}''&=\g_{gg}-r(u\inv-1)\g_q^{ss}-rv\g_g^{s}
+\smallfrac{d}{dt}\ln u.\cr}}
The colour-charge relation in the quark sector is thus again preserved
automatically. Starting for definiteness in a parton scheme
(such as, say the \QDIS\ scheme), where $F_2(x,t)= \langle e^2\rangle x
q(x,t)$ (and $\langle e^2\rangle$ is a numerical factor)
the functions $v$ and $\tilde v$ then give
the $F_2$
coefficient functions, since $q = (1-\tilde v)q''-vg''$. To
enforce the colour-charge  relation in the coefficient
functions we should
take
\eqn\cofuncc{\tilde v = rv+\hat v;\quad\hat v=-v_1\a}
(as is the case
in \MS\ scheme, for example~\CHad).
Similarly, the NLO terms violating the colour-charge relation at
NLO in the gluon sector can be removed  by choosing $rw-\tilde w$
appropriately.

The most general NLO
scheme change is thus parameterized by two parameters $u$ and $v$
(or more properly a
two-fold infinity of parameters, $u_n$ and $v_n$), plus a parameter
$\tilde v$ which is fixed requiring the coefficient functions to satisfy
the colour charge relation, and a parameter
$rw-\tilde w$ which can be used to impose the colour-charge relation
in NLO anomalous dimensions,
while the orthogonal combination $w+r\tilde w$ has no effect at all at
NLO. Given then parton distributions
in a particular parton scheme (such as \QDIS)
the parameter $u$ redefines the
normalization of the gluon, without affecting $F_2$ directly,
thus takes to different parton schemes (such as DIS or SDIS)
while $v$ moves to a different (generally non partonic) scheme
(such as  \MS).
%Imposing these two conditions, equations \twotfmns\ simplify to
%\eqn\uvtfmn{\eqalign{
%\g_{qg}''&=\g_{q}+(u\inv-1)\g_q^{ss}+v\g_g^{s},\cr
%\g_{gg}''&=\g_{g}-(u\inv-1)\g_q^{ss}-v\g_g^{s}
%+\smallfrac{d}{dt}\ln u,\cr
%\g_{qq}''&=r\g_{qg}''+\hat\g_q\qquad
%\g_{gq}''=r\g_{gg}''-\hat\g_q\cr
%}}
%so that the general form \nload\ is maintained, but now with
%$\hat\g_g''=-\hat\g_q$.

We can now extend these results to the physically relevant
double leading expansion, which treats the two large scales
on an equal footing (other expansion schemes~\Summing\ can be handled
in a similar way).
In this expansion, the matrix of singlet anomalous dimensions consists
of a large $x$ and a small $x$ contribution, $\g^L$ and $\g^S$, respectively:
\eqn\addl{\g=\g^L+\g^S;\qquad\g^L=\g^1+\g^2,\qquad\g^S=\g^{s}+\g^{ss},}
where $\g^1$ and $\g^2$ are the usual one and two loop anomalous
dimensions, while the leading and subleading singularities have the
form \singad, but now with $n_0=3$, so that both the one and the
two loop terms are suppressed to avoid double counting.
Scheme changes at LO are still effected by the matrix $U$  eq.~\udef\
(no nontrivial LO scheme change of $\g^L$ is possible). These transform
the singular anomalous dimensions according to eq.~\gamprime,~\gamprcc\
and are thus subject to the constraint eq.~\colchg. In addition, further
singular contributions are produced by the action of $U$ on $\g^L$, i.e.
$U\g^L U\inv$. These may be combined with the LO
transformation eq.~\gamprcc\ of $\g^S$ to give
\eqn\gampdl{\eqalign{\g_{qq}'&=\g_{qq}
+r(u\inv-1)\g_{qg},\cr
\g_{qg}'&=\g_{qg}+(u\inv-1)\g_{qg},\cr
\g_{gq}'&=\g_{gq}+(u-1)[(\g_{gq}-r\g_{gg}+r\g_{qq})
+r^2(u\inv-1)\g_{qg}]+r\smallfrac{d}{dt}\ln u,\cr
\g_{gg}'&=\g_{gg}-r(u\inv-1)\g_{qg}+\smallfrac{d}{dt}\ln u,\cr}}
independently of the split of $\g$ into $\g^L$ and $\g^S$.
Notice that all terms on the right hand side beyond the first are NLO
(because in particular $u\g^1=O(\a(\a/N)^n)$),
and that neither $\g^2$ nor $\g^s$ make any contribution to them.

Again the LO transformation redefines the normalization of the
gluon distribution without affecting the quark, which only transforms
under a NLO transformations. This now contains two components: a small $x$
NLO transformation of the form \vexp,~\vdef\ considered previously, plus
a standard
scheme changing $O(\a)$ function $V^L$, which vanishes as $N\to 0$:
\eqn\nlodlsc{
V=V^L+V^S;\qquad V^S = \a\sum_{n=0}^\infty V^s_{n+1}(\a/N)^n.}
Since $V^L$ has no effect on $\g^S$ at NLO, producing only terms
of $O(\a^2(\a/N)^n)$, we will not consider it any further here: its
only effect is to change the two loop anomalous dimension $\g^2$
in the usual way. The effect of the
remaining singular transformation $V^S$ is then very similar to that
already discussed in the small $x$ expansion:
$\g^S$ transforms according to the small $x$
 eq.~\gpp, while
the $O(\a)$ contribution to $V^S$ (which could in fact equivalently be viewed
as a contribution to $V^L$) also produces a NLO change of $\g^L$.
The general scheme transformation at NLO in the double leading expansion is
thus
\eqn\gamppdl{\g''=\g'+ \g^{s}_g\pmatrix{rv&v\cr
(rw-\tilde w-r\tilde v)&-rv\cr}+\a[V^s_1,\g^1],}
where here $\g^{s}_g$ is given by \singad\ with $n_0=1$
(i.e. including the one-loop contribution) $\tilde v$ was defined in
eq.~\cofuncc, and $\g'$ is the anomalous dimension transformed at
LO according to eq.~\gampdl.

We thus find again that, starting with a parton scheme, the parameter $v$
takes us to nonpartonic schemes, while the further parameters
$rw-\tilde w$ and $\tilde v$
may be fixed by requiring the color-charge relations in the
anomalous dimensions $\g^S$ and in the corresponding small $x$
coefficient functions, respectively. In comparison to standard scheme
change at large $x$~\FP\ there seems thus to be an additional
freedom, parametrized by $u$, of redefining the normalization of the gluon.
This redefinition by a LO function of $\alpha/N$
modifies the singular part of the NLO anomalous dimension (and appears
thus to be peculiar of small $x$ expansion schemes). The effect  of
this transformation can in practice be rather important.
Two (in some sense extremal) choices have
been considered in the literature. Taking $u(N,\a)=R(N,\a)\inv$, the
singular normalization factor for the gluon in \MS~\CHad, takes one
from the \QDIS\ scheme to the more singular \DIS\
scheme~\CHad. Because of the singular behavior of $R(N,\a)$
the anomalous dimensions $\g^{ss}_q$ in this scheme are substantially larger.
 Conversely, one may take $u=\g_{qg}/\g_{qg}^1$: this removes
the singular terms in the quark sector altogether, factoring
them in the initial gluon distribution (the `\SDIS' scheme\foot{
One could also consider a \SDIS$^\prime$
scheme \sdis, where  $u=\g_{qg}/(\g_{qg}^1+\g_{qg}^2)$, so that
$\g_q$ reduces to its two-loop expression}~\sdis).

The effect on the leading order relation between $F_L$ and the
gluon distribution is equally significant. In the double leading
expansion and \QDIS\ factorization scheme \ciaf\
$F_L=(C_L^1+h_L) g+O(\a^2)$, where $C_L^1$ is the usual
two loop longitudinal coefficient function, and
$h_L=\a\sum_1^\infty h_{n+1}(\a/N)^n$;
after the (LO) change of scheme
\eqn\fell{F_L = (C_L^1+h_L)(u\inv g' + r(u\inv -1) q')+ O(\a^2).}
The result is especially clear in
the small $x$ expansion, where the quark distribution is
subleading compared to the gluon \Summing:
the dominant contribution to $F_L$
is then equal to $F_L= h_L R g'+O(\a^2)$ in DIS~\CHad,
while in \SDIS\ $F_L= (h_L/h_2)c_q\a g'+O(\a^2)$ (since
$\g_q^1=c_q\a+O(\a^2)$).
Because $(h_L/h_2)=(1-\g_g^{s})/
(1+\smallfrac{3}{2}\g_g^{s}(1-\g_g^{s}))=1+O(\a/N)$, at
small $x$ $F_L$ will then differ from the gluon
distribution by a large factor. The leading order identification
of $F_L$ with the gluon may however
be restored by choosing a `GDIS' scheme in which $u=1+h_L/C_L^1$
so that $F_L=C_L^1\a g'$, so that in the small $x$ expansion
$F_L=c_q\a g'+O(\a^2)$ since $C_L^1=c_q\a+O(\a^2)$). In the \GDIS\
scheme $\g_q = (h_2/h_L)c_q\a +O(\a^2)$, so the quark anomalous dimensions
are less singular than in \QDIS\ or \DIS, but more singular
than in \SDIS.

This extra  freedom in the choice of factorization scheme at small $x$
can
however be constrained by requiring momentum conservation.
Combining the two transformations \gampdl\ and \gamppdl\ and imposing
the constraint of momentum conservation \momcon\ we get at $N=1$ the
two conditions
%\eqn\mom{-\beta_0\a^2\smallfrac{d}{d\a}u +
%\big(\g_g^{s}+\g_g^{ss}+r\g_q^{ss}-(r-1)v\g_g^{s}\big)u
%=(r-1)\g_q^{ss},}
%at $N=1$, the first of the momentum conservation conditions \momcon\
%will be satisfied at NLO. The condition
%\eqn\ccnlog{\hat\g_g+\hat\g_q+(u-1)(\hat\g_g+r\hat\g_q) + (rw-\tilde w -
%r\hat v)\g_g^{s} = 0.}
\eqnn\moma\eqnn\momb
$$\eqalignno{
-\beta_0\a^2\smallfrac{d}{d\a}u
+ \big(\g_g^{s}+\g_g^{ss}+r\g_q^{ss}
-(r-1)v\tilde\g_g^{s}\big)u
&=(r-1)\g_q^{ss},&\moma\cr
\hat\g_g+(u-1)\hat\g_g +
(rw-\tilde w -r\hat v)\tilde\g_g^{s} &=0,&\momb\cr}$$
where we used the crucial fact that momentum conservation at one
and two loops, i.e. $\g^1(1,\a)=\g^2(1,\a)=0$ naturally eliminates all
non-singular terms.
Notice that Eq.~\momb\ is equivalent to the
color-charge relation (in the subleading gluon sector),
which then does not have to be imposed separately, but
rather follows automatically.

Now, for each choice of $u$ eq.~\moma\ has a unique solution for
$v[u]$ [i.e. for the infinite set of coefficients $v_n(u_m)$],
while similarly \momb\ has a unique solution for $(rw-\tilde w)[u]$.
Conversely, since for a given $v$ eq.~\moma\ is a first order
differential equation for the function $u(1,\a)$ with boundary
condition $u(1,0)=1$ (compatible with \moma\ because
by construction $\gamma^s$ and $\g^{ss}$
contain no one loop terms),
it always has a unique solution $u[v]$, so the
functional $v[u]$ is invertible.\foot{While always true in QCD with
$N_c$ colours, since then $r=\half(1-1/N_c^2)$, in
supersymmetric Yang-Mills $r=1$,
so \moma\ fixes $u$ independently of $v$.}
It follows that eq.~\moma\ defines a
monotonic curve $u[v]$ in the two dimensional space of
schemes $(u,v)$ along which momentum is conserved: imposing momentum
conservation at NLO the gluon normalization is fixed uniquely
by the choice of $F_2$ coefficient functions.\foot{One may, of course,
 leave the line $v[u]$ while preserving momentum conservation
at NNLO: for example
one could perform a scheme change of the
conventional momentum conserving form $V=\bigl(
{{-r\hat v}\atop{r\hat v}}{{-\hat v}\atop{\hat v}}\bigr)$ (where
$\hat v$ is NLO). Under this change of scheme
$\g\to\g'$, where, if $\g$ satisfies the
colour-charge relation in both quark and gluon sectors
\eqn\gamppp{\g'=\g-\hat v(\g_g+\g_q)
\pmatrix{r&1\cr -r^2&-r\cr}+O(\a^2).}
If $\g$ satisfies momentum conservation, $(\g_q+\g_g)_{N=1}=0$
and the second term vanishes; however $v\g_q$, which must be kept
in order to get momentum conservation, is actually NNLO.
One would then be forced to include some NNLO contributions in a
NLO computations in order to preserve the momentum sum rule. }

In order to actually determine the curve $u[v]$, however,
one needs full knowledge of
the (unknown) subleading gluon anomalous dimension, i.e. of
$\g_g^{ss}$ and $\hat \g_g$
in a given scheme, say
\QDIS.
Indeed, the same argument which shows that for a given
$\g^S(1,\alpha)$ eq.~\moma\ determines
$v[u]$ or $u[v]$ also implies that for a given $v$ eq.~\moma\ determines
$u[\g^{ss}_g]$ (or $\g^{ss}_g[u]$): for every  $v$
and $\g^{ss}_g$ there exists
a $u$ which conserves momentum. Conversely,
for every pair of $u$ and $v$ momentum conservation
determines a unique two dimensional surface $\g^{ss}_g[u,v]$,
whose intersection with the plane $\g^{ss}_g=(\g^{ss}_g)_{\rm\QDIS}$ gives
back the curve $u[v]$.
Moving on the plane parametrized by $u$ and $v$ for fixed
$\g^{ss}_g$  yields  anomalous dimensions
which are equivalent up to a NLO change of scheme
(and only conserve momentum along the $u[v]$ curve); thus, it
explores the uncertainty related to the ignorance of NNLO
corrections. Different choices of $\g^{ss}_g$ then
explore the uncertainty related to the ignorance of the NLO gluon
anomalous dimension.

Momentum conservation constrains the uncertainty
in that it fixes one of the parameters in terms of the other two;
the overall uncertainty however is still necessarily NLO. Now,
given ${\g_g'}^{s}$ and ${\g_q'}^{ss}$ (from eq.~\gampdl\
for a certain choice of $u$ and $v$)
${\g_g'}^{ss}$ is determined algebraically (by eq.\momcon\ for $n>2$)
without the need for an explicit
computation~\Alf; it is thus convenient to
fix it thus, and vary $u$ for given $v$ in order to estimate
the corresponding NLO uncertainty~\Alf.
In practice, this uncertainty should only have minor effects on $F_2$, because
since $\g_{q}^{s}$ vanishes, $q$ evolves only at NLO, while although $g$
evolves at LO it only affects structure functions at NLO; the effect
of $\g_{g}^{ss}$ is then  subleading when compared to
$\g_{g}^{s}$, and thus effectively NNLO.

In conclusion, we have shown that momentum conservation can be enforced
in QCD evolution equations at NLO even when anomalous dimensions are
computed within an expansion scheme which sums up all leading and
subleading logs of both $1\over x$ and $Q^2$, such as the double
leading expansion \Summing. Within these expansion schemes there is a wider
freedom of choice of factorization scheme than in the usual (large $x$)
loop expansion of anomalous dimensions~\FP: besides the usual freedom
to perform a NLO scheme change which modifies
the $F_2$ coefficient functions, there is now also the possibility of
performing a LO scheme change which does not affect the $F_2$ coefficient
functions or the LO anomalous dimensions, but changes
the definition of the gluon distribution (or, equivalently,
the $F_L$ coefficient
function). The latter freedom is however fixed by the requirement of
momentum conservation. This is analogous to what happens in large
$x$ computations, where the momentum sum rule fixes the
normalization of the gluon; however at small $x$ the normalization
is actually given by a general LO function of $\a/N$,
rather than being just a number. Because the NLO $\g^{gg}$ and $\g^{gq}$
gluon anomalous dimensions are still unknown, this freedom cannot in practice
be pinned down, so that it is still necessary to vary
the gluon normalization within a given NLO factorization scheme, thereby
introducing a NLO uncertainty in the solution of the evolution
equations~\Alf. This uncertainty has fortunately only
sub-subleading effects on the structure
function $F_2$, however, due to the vanishing of the LO singularities
in the quark anomalous dimensions.

If $\g_g^{ss}$ were to be determined explicitly (for instance by
computing subleading corrections to the BFKL kernel in a particular
factorization scheme) it would then be possible use momentum
conservation to fix the definition of the gluon distribution,
thereby reducing the uncertainty in the solution of evolution
equations to a purely NNLO one. This could have significant  phenomenological
consequences in that it might substantially reduce the large scheme dependence
which is at present an intrinsic feature of perturbative computations at small
$x$ and relatively low $Q^2$~\ref\schem{S.~Forte and
R.~D.~Ball, CERN preprint CERN-TH/95-184, {\tt hep-ph/9507211}.}.

\bigskip
{\bf Acknowledgement}: We would like to thank Keith Ellis for
emphasising to us the importance of momentum conservation at small $x$.

\vfill\eject
\listrefs
\vfill\eject
%\listfigs
%\vfill\eject
\bye